\documentclass[amsmath,amssymb,amsfonts,aps,prb,twocolumn,superscriptaddress,footinbib]{revtex4-2}

\usepackage{graphicx}
\usepackage{dcolumn}
\usepackage{bm}
\usepackage[normalem]{ulem}
\usepackage{xcolor}
\usepackage{comment}
\usepackage{amsmath}
\usepackage{siunitx}
\usepackage{dsfont}
\usepackage{float}
\usepackage{placeins}
\usepackage[shortlabels]{enumitem}
\usepackage{xr}

\newcommand{\redmark}[1] {\color{red}#1\color{black}}

\newcommand{\up}{\uparrow}
\newcommand{\dn}{\downarrow}
\DeclareSIUnit\angstrom{\text{Å}}

\begin{document}

\title{Prediction of two-dimensional $\pi$-electron half-metallic ferrimagnets}

\date{\today}

\author{J. Phillips }
\affiliation{International Iberian Nanotechnology Laboratory (INL), Av. Mestre Jos\'e Veiga, 4715-330 Braga, Portugal}
\email{jan.phillips@inl.int}

\author{J. C. G. Henriques}
\affiliation{International Iberian Nanotechnology Laboratory (INL), Av. Mestre Jos\'e Veiga, 4715-330 Braga, Portugal}
\affiliation{Universidade de Santiago de Compostela, 15782 Santiago de Compostela, Spain}

\author{J. Fern\'andez-Rossier}
\altaffiliation[On permanent leave from ]{Departamento de F\'isica, Universidad de Alicante, 03690 San Vicente del Raspeig, Spain.}
\affiliation{International Iberian Nanotechnology Laboratory (INL), Av. Mestre Jos\'e Veiga, 4715-330 Braga, Portugal}
\email{joaquin.fernandez-rossier@inl.int}

\author{A. T. Costa}
\affiliation{Department of Physics, Center of Physics (CF-UM-UP), University of Minho, Campus of Gualtar, 4710-057, Braga, Portugal}
\affiliation{International Iberian Nanotechnology Laboratory (INL), Av. Mestre Jos\'e Veiga, 4715-330 Braga, Portugal}
\email{antonio.costa@inl.int \redmark{check}}

\begin{abstract} 
We propose a strategy to obtain conducting organic materials   with  fully spin-polarized Fermi surface, lying at a singular flat band, with  antiferromagnetically coupled magnetic  moments that reside in pi-orbitals of nanographenes. We consider a honeycomb crystal whose unit cell combines two different molecules with $S=1/2$: an Aza-3-Triangulene, a molecule with orbital degeneracy, and  a 2-Triangulene.
The analyzed system is half-metallic with a ferrimagnetic order, presenting a zero net total magnetic moment per unit cell. 
We combine density functional theory calculations with a  Hubbard model Hamiltonian to compute the magnetic interactions, the bands, the intrinsic Anomalous Hall effect, and the collective spin excitations. We obtain very large intermolecular exchange couplings, in the range of 59 meV.  Based on the spin excitation dispersion, we estimate  thermal stability  in the range of 100 Kelvin.  When the magnetization is off-plane,  intrinsic spin orbit coupling in graphene opens up a topological gap that, despite being very small, leads to a quantized Hall conductance in the tens of mK range, but is thermally smeared above 1 Kelvin.

\end{abstract}
\maketitle

\section{Introduction}
The search for materials exhibiting spin-split electronic bands and vanishing net magnetization is an extremely active area of research\cite{Yuan2020, Hayami2020, Yuan2021, Dong2025}. Spin splitting is a key figure of merit for many spintronic applications, including tunnel magnetoresistance\cite{jungwirth2016antiferromagnetic, Dong2022, Smejkal2022TunnelingMagneto}. In conventional spintronic devices, spin polarization is typically provided by ferromagnetic layers; however, these generate stray magnetic fields that couple to other circuit elements and are detrimental for sensing and memory applications. In this context, altermagnetic spintronics has recently emerged as a particularly active field, as altermagnets exhibit spin-split bands along specific crystallographic directions while maintaining zero net magnetization\cite{jungwirth2022alter1, jungwirth2022alter2}.

Half-metallic fully compensated ferrimagnets\cite{deGroot1983halfmetal,deGroot1991compensatedHM,pickett1998compensatedHM,galanakis2002halfmetal} constitute another class of materials sharing several of the properties that make altermagnets fundamentally appealing. Previous theoretical studies propose 2D models for fully compensated ferrimagnets \cite{liu20205FullyCompFerri2D} and organic platforms \cite{Kawamura2024organicFerri}. These systems display complete spin polarization of the electronic states together with zero total magnetization. The magnetic moments of the distinct sublattices are antiparallel and cancel macroscopically, while the electronic structure remains half-metallic, with a band gap in one spin channel and metallic behavior in the other. This combination enables fully spin-polarized charge transport in the absence of stray magnetic fields, distinguishing these materials from conventional ferromagnetic half-metals. As a result, half-metallic fully compensated ferrimagnets are of strong fundamental interest for elucidating the interplay between electronic structure and magnetic order and are highly attractive for spintronic applications\cite{vzelezny2018spin,finley2020spintronicsHM}.

Here, we propose a route to engineering half-metallic fully compensated ferrimagnetism in triangulene crystals. Triangulenes are a prototypical class of open-shell nanographene structures that serve as experimentally realizable building blocks for constructing larger molecular architectures with tailored magnetic properties\cite{zhou20, mishra2019b, li2020uncovering, mishra2020, mishra2020topological, Ortiz2020, turco23, mishra2021large, mishra21, hieulle2021, wang2022, cheng2022surface, yu2025metal, yu23dimer, Yu2024physics, anindya2022, Ortiz2026}. Advances in on-surface synthesis and ultra-high-vacuum experimental techniques \cite{cai2010, moreno2018, Song2021} have enabled the fabrication of triangulenes with different sizes\cite{pavlivcek2017, su2019, mishra2019b, mishra2021b, turco23}, as well as dimers\cite{mishra2020}, rings\cite{mishra21, hieulle2021}, chains\cite{mishra21}, and two-dimensional lattices\cite{delgado23, Pawlak2025}.

Our starting point is the recent observation that non-centrosymmetric triangulene crystals, with a unit cell composed of a [2]triangulene (phenalenyl) and a [3]trianguelene, as shown in Fig.~\ref{fig:struct}(a), are uncompensated ferrimagnetic insulators with spin-split bands\cite{catarina2023} (see Fig.~\ref{fig:struct}(b),(c). In this system, the [3]triangulene carries spin $S=1$, while the [2]triangulene carries $S=1/2$. The resulting band structure is highly unconventional: in the majority (minority) spin channel, shown in red (blue), a flat band merges with a dispersive band at the $\Gamma$ point, forming a singular flat band that is fully occupied (empty). This flat band originates from a zero mode localized on the [3]triangulene molecule.  

In order to make this structure conducting, we explore the addition of a substitutional  nitrogen atom into the central carbon site of the [3]triangulene. This introduces an additional minority-spin electron, thereby reducing the magnetic moment of the [3]triangulene, promoting compensation of the total magnetization within the unit cell, and shifting the Fermi level into the conduction band. Both density functional theory calculations, performed using Quantum Espresso at the generalized gradient approximation level with PBE functionals, and mean-field Hubbard model calculations confirm this scenario. The resulting structure simultaneously exhibits three key features: a vanishing total magnetic moment (Fig.~\ref{fig:struct}(d)), a Fermi energy located immediately below the singular flat band at the top of a dispersive band (Fig.~\ref{fig:struct}(e),(f), and a very large density of states at the Fermi level that is fully spin polarized.

The rest of the manuscript is organized as follows. In section \ref{sec:methods} we describe details of the density functional theory (DFT) and mean-field Hubbard (MFH) calculations used to obtain our results. Then, follows section  \ref{sec:electronic}, where we discuss the electronic structure of the crystal and the possibility of Anomalous Hall effect in this system. We discuss the spin excitations of the [2,A3]triangulene crystal in section \ref{sec:magnons}, obtained through the random phase approximation (RPA) method. In section \ref{sec:stm} we analyze the elastic and inelastic scattering contributions that would be present in hypothetical scanning tunneling microscopy (STM) measurements. Finally, in section \ref{sec:conclusions}, we discuss the conclusions of our work.

\section{Methods\label{sec:methods}}

\subsection{DFT-based calculations}
We carry out two types of electronic structure calculations: collinear MFH model and DFT) calculations. DFT results were obtained using the {\sc QUANTUM ESPRESSO} package \cite{QE-2017, QE-2009, doi:10.1063/5.0005082}. We used norm-conserving pseudopotentials \cite{PhysRevLett.43.1494, PhysRevB.43.1993, van_Setten_2018, Hamann2013, garrity2014pseudopotentials, doi:10.1126/science.aad3000}. The plane-wave energy cutoff of was converged to a value of 70 Ry for the wave functions and 280 Ry for the charge density. The Brillouin zone integrals for the relaxations were performed in a converged 15$\times$15$\times$1 k-point grid with a gaussian smearing of 0.001 Ry. In-plane lattice parameters and internal atomic positions were relaxed for every doped structure. The crystals include a converged 9 \AA \ vacuum for the z-axis to avoid inter-molecular interactions.

\subsection{Mean field Hubbard calculations}
For the MFH calculations our starting point is a tight binding Hamiltonian accounting only for the $p_z$ Carbon orbitals, with first, $t_1 = -2.7$ eV, and third, $t_3 = t_1/10$, neighbor hoppings, to which a Hubbard interaction, $U = |t_1|$, is added. On top of this, other single particle interactions, like the Kane-Mele interaction \cite{kane2005quantum,min06} we consider later in the text, can be trivially introduced. The mean field decoupling is performed on the Hubbard part as:
\begin{equation}
    H_U \approx \sum_i U_i \left[n_{i\uparrow}\langle n_{i\downarrow} \rangle +\langle n_{i\uparrow}\rangle  n_{i\downarrow}-\langle n_{i\uparrow}\rangle \langle n_{i\downarrow}\rangle\right]
    \label{eq:HubbardMF}
\end{equation}
where $n_{i\sigma}$ is the occupation operator of the orbital in site $i$ with spin $\sigma = \up/\downarrow$. The expectation values are taken over the MFH ground state. The broken symmetry equilibrium configuration is obtained following the standard iterative self-consistent approach, starting from an initial guess for the ground state, until convergence is reached \cite{fernandez07,fernandez08,Culchac2011,catarina2023,henriques2024}.

\subsection{Calculation of the spin susceptibility using RPA}

The energies and lifetimes of the magnons with $S^z=-1$ and $S^z=1$ were obtained from the dynamic transverse spin susceptibilities, calculated within the RPA,
\begin{eqnarray}
    \chi^{+-}_{ll'}(t)\equiv -i\theta(t)\left\langle\left[S^+_l(t),S_{l'}^-(0)\right]\right\rangle,\nonumber\\
    \chi^{-+}_{ll'}(t)\equiv -i\theta(t)\left\langle\left[S^-_l(t),S_{l'}^+(0)\right]\right\rangle,
    \label{eq:chidefinition}
\end{eqnarray}
where $S^+_l\equiv a^\dagger_{l\up}a_{l\dn}$, $S^-_l\equiv a^\dagger_{l\dn}a_{l\up}$, $l$ is a site index, $\langle\cdot\rangle$ is a thermodynamic average and $\theta(t)$ is the Heaviside unit step function.
The magnon spectral densities are defined as the imaginary part of the frequency-domain susceptibilities,
\begin{equation}
    \chi^{\perp}_{ll'}(\omega) = \int_{-\infty}^\infty e^{i\omega t} \chi^{\perp}_{ll'}(t)dt,
\end{equation}
where $\perp$ stands for either $+-$ or $-+$, and $l,l'$ are atomic site indices.
The main result of the RPA\footnote{See Supplemental Material at [URL] for more details about the RPA calculations in the reference list.} decoupling scheme is the relationship between the
``interacting'' $\chi$ and mean-field $\tilde{\chi}$ susceptibilities,
\begin{equation}
     \chi^\perp_{ll'}(\omega;\vec{q}) =  \tilde{\chi}^\perp_{ll'}(\omega;\vec{q}) + \sum_m \tilde{\chi}^\perp_{lm}(\omega;\vec{q})U_{m}\chi^\perp_{ml'}(\omega;\vec{q}) .
\end{equation}

\section{Electronic properties\label{sec:electronic}}

\begin{figure}
    \centering
    \begin{tabular}{cc}
       \includegraphics[width=\linewidth]{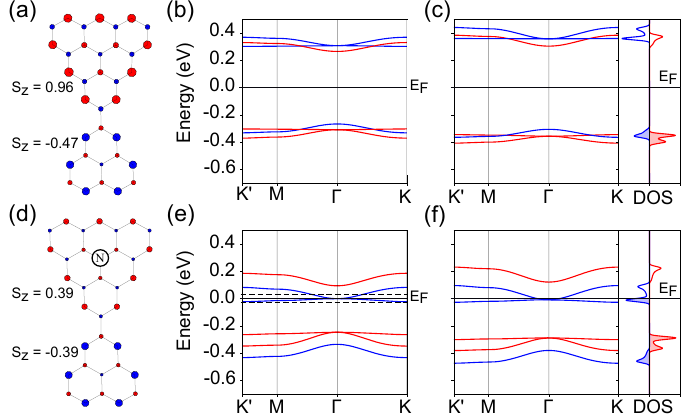} &

    \end{tabular}
    \caption{DFT computed magnetic profiles, electronic bandstructure and DOS (eV$^{-1}$)  for the [2,3]triangulene (a),(b),(c), and for the [2,A3]triangulene (d),(e),(f). Spin up (down) is represented in red (blue) in the bandstructures computed via MFH (b),(e) and for the DFT magnetic profiles, bandstructures and DOS (c),(f). Solid line at 0 energy represents the Fermi level (E$_F$). Horizontal dashed lines in e), located at $\pm 25 meV$, represent the values of the scanned energies used for computing LDOS. 
    }
    \label{fig:struct}
\end{figure}

\subsection{Energy bands and spin-resolved density of states}

Our calculations for the resulting [2,A3]triangulene crystal confirm that the ground state retains the antiferromagnetic coupling between individual triangulenes, which is already present in the undoped crystal (Fig.~\ref{fig:struct}(a) due to intermolecular exchange interactions. The structural DFT relaxation of lattice parameters and internal atomic positions shows a small reconstruction of the atomic distances surrounding the central [3]triangulene site compared to the undoped system. The atomic distances from this central D$_3$h symmetric site decrease from 1.42 \AA \ to 1.40 \AA \ when substituting the C atom for N, but preserving the original symmetry of the nitrogen free case. The symmetry reduction from the D$_3$h to the C$_2$v reported for the monomer in \cite{wang2022}, is absent, and we ascribe this to the lifting of energy degeneracy in the frontier orbitals due to intermolecular hybridization, as previously studied in \cite{henriques2024}. Therefore, Jahn-Teller distortion is only present in the monomer. 
The 59-meV AF–FM energy difference indicates a large magnetic exchange scale. Consequently, the [2,A3]triangulene crystal exhibits a total net spin of $S_z=0$ in its ground state and thus realizes a fully compensated ferrimagnet.

We have verified that  doping the central site with phosphorus or boron, instead of N, produces very similar results (see Fig. S2). Doping with a heavier atom such as P is equivalent to N substitution, as it introduces an extra electron into the system, shifting the Fermi level upward toward the Dirac point. Interestingly, using B instead of N or P, also leads to a half-metallic, fully compensated ferrimagnet. In this case, B  removes an electron, lowering the Fermi level toward the Dirac point of the spin-up channel. In both cases, the Fermi level lies in close proximity to the flat bands, and the antiferromagnetic configuration remains the ground state. This configuration is energetically favored over the ferromagnetic one by approximately 81 meV for the P-doped system and 46 meV for the B-doped system, suggesting stability at high temperature. Together, these results reinforce those obtained for the [2,A3]triangulene crystal and demonstrate the robustness of our proposed strategy for realizing half-metallic fully compensated ferrimagnets.

\subsection{Anomalous Hall effect}

\begin{figure}
    \centering
    \includegraphics[width=\linewidth]{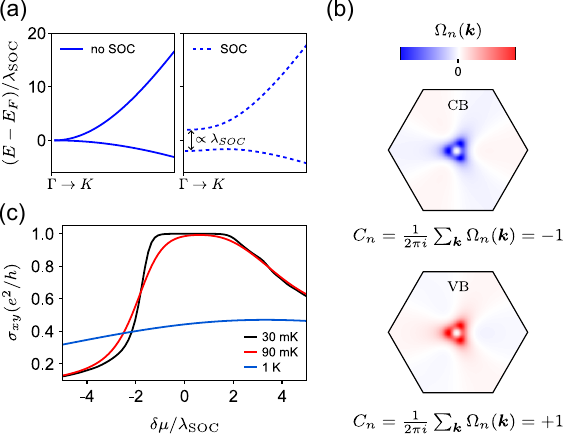}
    \caption{(a) Gap opening at the $\Gamma$ point in the presence of intrinsic SOC. The size of the gap is proportional to the strength of SOC i.e. $\lambda_\textrm{SOC}$. (b) Berry curvature, $\Omega_n (\boldsymbol{k}$), in the first Brillouin zone for the the valence band (VB) and conduction band (CB). (c) Zero frequency Hall conductivity, as a function of the chemical potential detuning, $\delta\mu = \mu - E_F$, at three different temperatures. A value of $\lambda_{SOC} = 45 \mu$ eV was used in the simulations.}  
    \label{fig:ahe}
\end{figure}

Having engineered the location of the Fermi level, we now discuss a remarkable property which arises from it: a finite Anomalous Hall Effect (AHE) despite the vanishing net magnetization in the crystal's unit-cell. The system we present complements recent studies on altermagnets and non-coplanar antiferromagnetic phases which also show finite AHE \cite{vsmejkal2022anomalous}. 

By analogy with the px-py honeycomb model \cite{wu2007flat}, that also features singular flat bands at the $\Gamma$ point, we expect that  the presence of spin-orbit coupling (SOC)  will open up a small gap at the $\Gamma$ point \cite{zhang2014honeycomb}, thus lifting the degeneracy between the dispersive and the nearly-flat bands. To capture this, we introduce a Kane-Mele \cite{kane2005quantum,min06} intrinsic SOC term in the model Hamiltonian: 
\begin{align}
    H_\textrm{KM}= i \lambda_{\textrm{SOC}} \sum_{\langle\langle i,j \rangle\rangle} \sum_{\sigma} \sigma \nu_{ij} c^\dagger_{i,\sigma} c_{j,\sigma} + \textrm{h.c.}
    \label{eq:KMHamilt}
\end{align}
where $\lambda_{\textrm{SOC}}$ quantifies the strength of intrinsic SOC ($\lambda_\textrm{SOC} = 45\mu$eV in what follows), the first sum runs over second neighbors sites $(i,j)$, the second over the spin component $\sigma = \up,\downarrow$ and $\nu_{ij} = \pm 1 =  \textrm{sign}[\hat{z}\cdot(\boldsymbol{r}_i \times \boldsymbol{r}_j)]$, depending on whether the electron makes a left (right) turn to go from site $i$ at $\boldsymbol{r}_i$ to site $j$ at $\boldsymbol{r}_j$. 

In Fig. \ref{fig:ahe}(a) we show that, as expected, in the presence of SOC a small gap opens at the $\Gamma$ point, with a magnitude proportional to $\lambda_\textrm{SOC}$. In this gapped phase the system becomes a topologically non-trivial Chern insulator. The bands around the Fermi level carry a finite Chern number \cite{fukui2005chern, hasan2010colloquium}, with $C_n = +1$ for the valence band (VB) and $C_n = -1$ for the conduction band (CB), as depicted in Fig. \ref{fig:ahe}(b). As $\lambda_\textrm{SOC}$ decreases, the hot-spots of Berry curvature \cite{berry1984quantal, hasan2010colloquium}, $\Omega_n(\boldsymbol{k})$, seen in Fig. \ref{fig:ahe}(b), approach the $\Gamma$ point and perfectly cancel out in the $\lambda_\textrm{SOC} \rightarrow 0$ limit, when the two bands touch, where the Chern-insulating phase ceases to be defined.

A direct consequence of the finite Berry curvature is a finite value of the intrinsic Hall conductivity, $\sigma_{xy}$, defined as \cite{fukui2005chern, hasan2010colloquium, vsmejkal2022anomalous}:
\begin{align}
    \sigma_{xy} = -\frac{e^2}{2 \pi h} \sum_{n,\boldsymbol{k}} \Omega_n ( \boldsymbol{k} ) f(E_{n} (\boldsymbol{k}),\mu,T), \label{eq: sigma_xy}
\end{align}
where the sum over $n$ runs over all bands of the system, $\boldsymbol{k}$ is in the first Brillouin zone and $f(E_{n}(\boldsymbol{k}),\mu,T)$ is the Fermi-Dirac distribution for a state with energy $E_{n}(\boldsymbol{k})$, chemical potential $\mu$ and temperature $T$. 

In Fig. \ref{fig:ahe}(c) we show the intrinsic Hall conductivity, as given in Eq. (\ref{eq: sigma_xy}), as a function of the chemical potential detuning, $\delta\mu = \mu - E_F$ at three different temperatures. At the lowest temperature, $T = 30$mK, we find a quantized Hall conductivity when $\delta\mu$ is small. In such a limit, the bands below the chemical potential are fully occupied, and the ones above it are fully empty; this reduces Eq. (\ref{eq: sigma_xy}) to a sum over Chern numbers. We stress that this sum is only finite due to the carefully engineered location of $E_F$, and it would vanish in the nitrogen-free case.
As $|\delta \mu|$ keeps increasing, $\mu$ eventually penetrates the VB or the CB bands and $\sigma_{xy}$ stops being quantized as bands become partially filled. 
Repeating the same analysis for higher temperatures, a similar behavior is found, but the region where $\sigma_{xy}$ is quantized decreases (and may even disappear), due to thermal occupation of high energy bands.  This thermal occupation leads to a partial cancellation of the Berry curvatures of the VB and CB, taking $\sigma_{xy}$ away from $e^2/h$, as we see in the simulation with $T = 1$K. 
Increasing $\lambda_{\textrm{SOC}}$, which could be achieve through doping with heavier atoms, for example, would increase the gap between VB and CB, leading to a wider energy window and larger temperature range where a quantized $\sigma_{xy}$ could be measured.
In Ref. \cite{chang2013experimental}, a Hall conductivity as small as $\sigma_{xy} = 0.2e^2/h$ was measured at $T = 30$mK, showing that it should be possible to probe the AHE in the system presently discussed.

\section{Spin excitations\label{sec:magnons}}

\begin{figure}
    \centering
    \includegraphics[width=\linewidth]{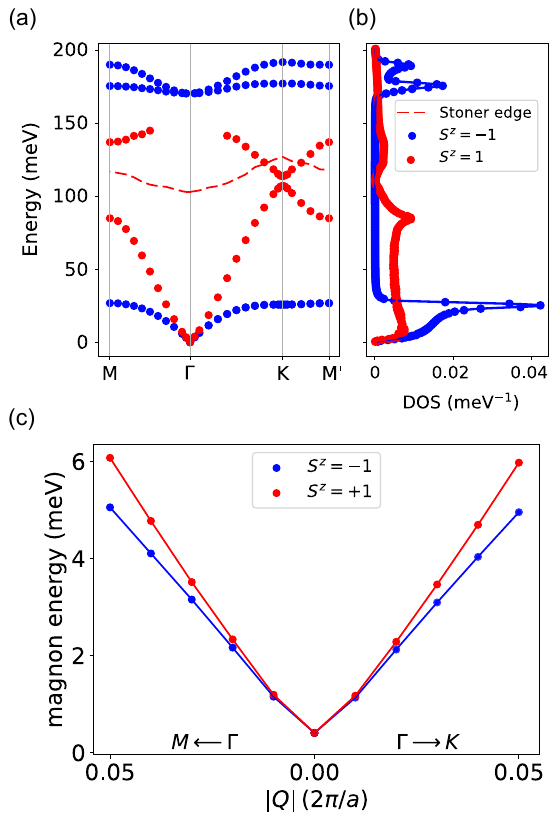}
    \caption{(a) Magnon energies for selected wave vectors extracted from the fermionic model, for the two magnon flavors $S^z=-1$ (blue) and $S^z=1$ (red). The red dashed line marks the lower energy boundary of the $S^z=1$ Stoner continuum. Notice the large magnon energies at the edges of the BZ, compatible with the large exchange estimated from the DFT calculation. (b) Total magnon density of states for $S^z=-1$ (blue) and $S^z=1$ (red). The solid lines are guides to the eye. (c) Magnon energies as functions of wave vector in the presence of effective SOC (Eq.~\ref{eq:KMHamilt}) with $\lambda_\mathrm{SOC}=45\mu\mathrm{eV}$. The anisotropy gap at $\Gamma$ is $\Delta_\Gamma\approx 0.4$~meV.}
    \label{fig:magnons}
\end{figure}

A half-metallic fully compensated ferrimagnet is also peculiar from the point of view of its spin excitations. The metallic character of the electronic structure suggests strong electron-magnon coupling, which provides means of manipulating the magnetic state electrically but, on the other hand, may significantly reduce magnon lifetime through Landau damping~\cite{Barbosa2001,MCosta2020,henriques2024,Costa2025}.
The Néel-type magnetic ordering implies that magnons come in two flavors, distinguished by their $S^z=\pm 1$ quantum number. However, in sharp contrast to antiferromagnets, due to the spin polarization of the electronic bands $S^z=\pm 1$ magnons will have different energies, which depend on the the electronic bands in a non-trivial way. To explore those properties we calculated the magnon spectrum, which is associated with the transverse components of the dynamic spin susceptibility tensor $\chi^\perp(\omega,\vec{q})$, within the RPA)~\cite{Barbosa2001,Costa2025,henriques2023}. Magnon energies $\hbar\omega$ for a fixed wave vector $\vec{q}$ are associated with the positions of the peaks of $-\mathrm{Im}\chi^\perp(\omega,\vec{q})$ as $\omega$ is scanned. \
A dispersion relation for magnons along high symmetry lines in the Brillouin zone is shown in Fig.~\ref{fig:magnons}(a). We find two magnon branches with distinct energies for each wave vector, corresponding to $S^z=\pm 1$. The modes are degenerate only at the $\Gamma$ point where their energy vanish if spin-orbit coupling is neglected, complying with the Goldstone theorem. 
The energy differences between the two modes are much larger than those found in altermagnets, for instance~\cite{Costa2025}. This can be qualitatively understood in terms of the large spin polarization of the electronic bands, see Fig.~\ref{fig:struct}. 

The two lowest energy branches in Fig.~\ref{fig:magnons}(a) correspond to modes in which the whole magnetization of each triangulene behave as a single, monolithic, spin (see supplementary material for a detailed discussion). This system also has higher energy $S^z=-1$ magnons that correspond to ``internal'' excitation modes of the individual aza-triangulene (the blue circles in the energy range of 150-200 meV). We had already encountered this kind of magnons in a half-metallic ferromagnetic system~\cite{henriques2024}. Here, however, the internal modes have energy below the edge of the Stoner continuum with $S^z=-1$ ($\sim 250$~meV), which protects them from Landau damping. Finally, we also find a second branch of $S^z=1$ magnons, with energies in the range 110-150~meV. These, however, overlap with the $S^z=1$ Stoner continuum except for a small pocket around the $K$ point in the Brillouin zone. The missing energy points in the region around the $\Gamma$ point correspond to overdamped $S^z=1$ magnons. This analysis highlights the fact that it is impossible to describe the spin dynamics of this material with a spin-only model where we assign a rigid spin per triangulene. Doping one of the triangulenes with nitrogen brings orbital degrees of freedom into play, which drastically affect the spin dynamics.  

A metallic magnetic system in which most magnon modes live below the Stoner continuum is very unusual. This fact, as discussed previously, protect the magnons from Landau damping, thus blocking the main channel for magnon decay in metallic magnets. The remaining sources of magnon damping are either magnon-magnon interactions, which are expected to contribute little at low temperatures, and spin-orbit coupling mediated decay, which is negligible in carbon-based materials. Thus, this material seats at the sweet spot where magnons can be created electrically but they have a extremely narrow intrinsic linewidth,  making it a prime candidate for use in magnonic devices~\cite{Fert2024RevModPhys}.

\subsection{Estimation of the transition temperature}
To obtain a rough estimate of the transition temperature associated with the magnetic order, 
we performed calculations of the magnon energies close to the $\Gamma$-point in the presence 
of an effective spin-orbit coupling~\cite{Costa2010SOC,Costa2020CrI3} described by a Kane-Mele term~\cite{kane2005quantum}, Eq. (\ref{eq:KMHamilt}), like in the description of the AHE above.
The term opens up an anisotropy gap that regularizes the magnon occupancies at low temperatures
and circumvents the prohibition of long-range order imposed by the Mermin-Wagner theorem. For the the value of the Kane-Mele parameter used in the calculation of the AHE ($\lambda_\mathrm{SOC}=0.45\mu\mathrm{eV}$), we find numerically that the magnon anisotropy gap
is $\Delta_\Gamma\sim 0.4\,\mathrm{meV}$. We note that in  antiferromagnets, the magnon anisotropy gap is controlled by the geometric mean of the exchange and the magnetic anisotropy.  The large magnitude of intermolecular exchange explains  why the magnon gap comes out larger than the spin-orbit coupling\cite{kittel51}. A plot of the dispersion relation of the two lowest magnon branches around the $\Gamma$-point is shown in Fig.~\ref{fig:magnons}(c).  Despite its very small value, this gap is 
enough to stabilize the magnetic order at temperatures up to several tens of Kelvin. 
The estimate for the transition temperature can be obtained from the renormalized spin 
wave theory expression for the thermal depletion temperature for each sublattice~\cite{TahirKheli1962,Callen1963,Lee1967},
\begin{equation}
    \frac{1}{T^{(0)}_s} = \frac{1}{N}\sum_{\vec{k}}\left( \frac{\xi^s_1(\vec{k})}{\omega_1(\vec{k})} + \frac{ \xi^s_2(\vec{k}) }{ \omega_2(\vec{k}) } \right), 
\end{equation}
where $\omega_i(\vec{k})$ are the magnon dispersion relations for the two lowest magnon branches, 
including the anisotropy gap, and $\xi^s_i(\vec{k})$ are the Bogoliubov weights at sublattice $s$ for mode $i$, arising from linear spin wave theory applied to an effective spin Hamiltonian~\cite{Pires2021Book}. These are the temperatures at which the magnetization of each sublattice go to zero due to the thermal population of magnon states. By fitting the two lowest 
magnon branches to a spin model on a honeycomb lattice and evaluating the expression above for 
each sublattice we obtain $T^{(0)}_{A}\approx 79\,\mathrm{K}$ for one sublattice and 
$T^{(0)}_{B}\approx 130\,\mathrm{K}$ for the other. This indicates transition temperatures in the ballpark of $10^2$~K.

\section{STM spectroscopy \label{sec:stm} }

\begin{figure*}
    \centering
    \includegraphics[width=\linewidth]{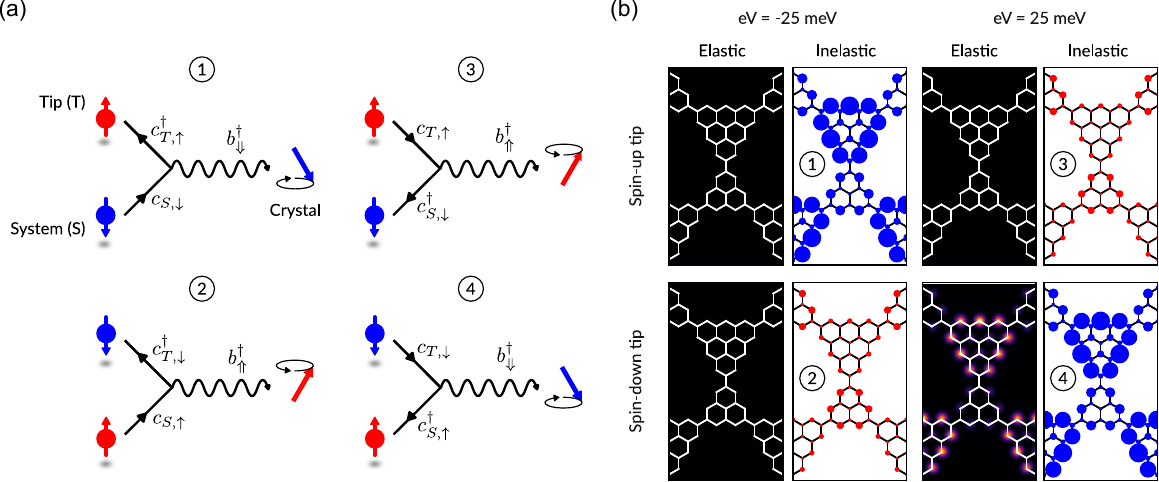}
    \caption{\textbf{Detecting magnons with inelastic scanning tunneling spectroscopy}. The left panel (a) shows schematic depictions of the inelastic tunneling processes that can lead to a feature associated with the emission of a magnon in the differential conductance of an ISTS experiment. The right panel (b) shows local electron (elastic) and magnon (inelastic) local density of states maps for two values of the tip-sample bias: at $eV=-25\textrm{ meV}$ elastic tunneling from the triangulene crystal is suppressed for both spin polarizations of the tip due to the absence of available electronic states; inelastic tunneling is possible by creating a magnon of either $S^z=\pm 1$. At $eV=25\textrm{ meV}$, elastic tunneling is suppressed for tip polarization parallel to that of the individual aza-triangulene because the electronic band close to the Fermi level is polarized in the opposite direction. Inelastic tunneling is possible for both spin polarizations of the tip, each by exciting a magnon with the same energy polarization as the tip.   }
    \label{fig:LDOS}
\end{figure*}

 Having discussed the spin excitations, the question of how to probe them in the ferrimagnetic phase using scanning tunneling spectroscopy (STS) can be addressed. Both elastic and inelastic tunneling processes  contribute to the conductance and provide valuable information\cite{Ortiz2020}.  The elastic contribution provides information about the electronic density of states,  and the inelastic conductance is roughly related to the spin excitation density of states.  We assume the [2,A3]triangulene  crystal is deposited on top of a conductor, chosen so that there is no charge transfer. Importantly,  depending on the spin of the tunneling electron and the voltage range, the electrons leaving the STM tip can tunnel either into electronic states of the [2,A3]triangulene  crystal or into the underlying conducting substrate. The strength of these two processes is very different, on account of the exponential decay of the tunneling amplitude.
 
 We assume that the tip is grounded, so that voltage drop occurs entirely on the surface electrode. Throughout, we define the bias as $V=V_s-V_t$, with $e>0$, so that $V>0$ ($V<0$) corresponds to electron tunneling from the tip to the sample (from the sample to the tip).  The rest of the discussion assumes that the ground state configuration has the A3T with positive ($\uparrow$) spin polarization and we discuss the contributions of  electron tunneling from/to the tip in each spin channel separately, assuming the same quantization axis for tip and sample. 

Our intention is not to attempt to carry out a quantitative calculation of the conductance. Instead,  the following discussion provides an heuristic description, based  on the Tersoff-Hamman approximation~\cite{TersoffHamman} for the differential conductance in a scanning tunneling microscope,
 \begin{equation}
     \frac{dI}{dV}\propto n^\up_\mathrm{tip}n^\up_\mathrm{sample}(E_F+eV)+n^\dn_\mathrm{tip}n^\dn_\mathrm{sample}(E_F+eV),
 \end{equation}
 where $n^\sigma$ is the density of states for spin channel $\sigma$ and $V$ is the tip-sample bias voltage.
 Formally, the tunnel conductance can be related to a convolution of the spectral function of the tip and the surface\cite{Mahan2000}. In the independent electron approximation, the spectral function of the surface is proportional to the density of states. The spin excitations of the sample renormalize its density of states~\cite{Schweflinghaus2014}, 
 leading to signatures in the $dI/dV$ spectrum which are roughly proportional to the density of magnon states, 
 shown in Fig.~\ref{fig:magnons}(b).  

 To illustrate the fact that this system allows observation of the inelastic signals without elastic contributions, an appropriate choice of bias and tip spin polarization must be made. Our choice of bias ($V=\pm 25$meV) was such that it coincided with the peak of the $S_z=-1$ (blue) magnon, as seen in Fig.~\ref{fig:magnons}(b). We now break down the tunneling processes in four groups, depending both on the sign of the bias and the spin of the electron that comes out/arrives to the tip: 

\begin{enumerate}

\item \textit{$|e|V =-25$ meV, electrons tunnel to the tip. \newline Initial spin: $\sigma=\uparrow$ (red):}\newline At this bias there are no electronic states in the crystal (for either spin), as shown in Fig. \ref{fig:struct}. Hence, elastic conductance signal can only appear due to tunneling directly from the conducting substrate. This contributions should be very weak, on account of the exponential decay of tunneling amplitude.  In contrast, a spin $\downarrow$ electron can travel from the substrate to the tip, flipping its spin and exciting a $S_z=-1$ magnon in the crystal (see Fig. \ref{fig:magnons}(a)). Since there is a high magnon DOS at this energy this inelastic process is expected to produce a strong signal (Fig. \ref{fig:magnons}(b)).

\item \textit{$|e|V =-25$ meV, electrons tunnel to the tip. \newline Initial spin: $\sigma=\downarrow$ (blue):}\newline There are still no electronic states in the [2,A3]triangulene crystal at this bias, therefore spin $\downarrow$ electrons come from the substrate and the elastic contribution is again very week. Nevertheless, spin $\uparrow$ electrons originating from the substrate can tunnel to the tip, flip its spin and produce a $S_z=+1$ magnon in the crystal. These spin excitations present a lower DOS than the previous case (Fig. \ref{fig:magnons}),and therefore should originate a weaker signal. 

\item \textit{$|e|V =+25$ meV,  electrons tunnel from the tip. Initial spin: $\sigma=\uparrow$ (red):} \newline  The elastic process is suppressed, because there are no electronic $\uparrow$ [2,A3]triangulene states close to the Fermi surface (see Figure \ref{fig:struct}), and the final state must therefore reside in the substrate.  In contrast, the inelastic process is allowed, where the $\uparrow$ electron from the tip tunnels to the substrate with spin flip, leading to the emission of a $S_z=+1$ magnon in the crystal.

\item\textit{$|e|V =+25$ meV, electrons tunnel from the tip. Initial spin: $\sigma=\downarrow$ (blue):}\newline Now the elastic tunneling to the [2,A3]triangulene  crystal is allowed, and a strong signal should be measured. Superimposed with the elastic contribution, signal coming from the inelastic process where a $S_z=-1$ magnon is emitted in the crystal, should also appear (Fig. \ref{fig:magnons}).
\end{enumerate}

These results are summarized in Table \ref{tab:tunneling_channels}.    There are two important  highlights. First, if the STM tip has a large (ideally complete) spin polarization, it is possible to {\em selectively} inject spin excitations with a given sign of $S_z$, linked to the sign of the bias.  This is reminiscent of the 
current-driven  emission of spin waves\cite{Berger96} and may be used to implement 
current induced spin torque\cite{Slonczewski96}, whose intensity could be modulated by lateral displacement of the STM tip with atomic resolution \cite{delgado10}.  Second,  also for spin-polarized tips,  we expect a large tunnel magnetoresistance (TMR), for both bias polarities,  on account of the full spin polarization of the Fermi surface of the [2,A3]triangulene  crystal:  if either the tip polarization of the magnetic state of the crystal are reversed, there will be a huge  variation of the tunnel conductance, as long as the bias remains in the region  where the single-particle DOS is fully polarized.

\begin{table}[t]
\centering
\caption{Summary of the elastic and one-magnon tunneling channels.
``Strong'' (``weak'') denotes a process involving electronic states
of the [2,A3]triangulene  crystal (underlying conducting substrate), respectively.
$S_z$ denotes the angular momentum of the emitted magnon.}
\label{tab:tunneling_channels}
\begin{tabular}{c c c c c}
\hline\hline
$\#$ & Bias (meV) & Tip spin & Elastic & Inelastic \\
\hline
1
& $eV=-25$
& $\uparrow$
& Weak
& $S_z=-1$ 
\\
2
& $eV=-25$
& $\downarrow$
& Weak
& $S_z=+1$ 
\\

& $eV=+25$
& $\uparrow$
& Weak
& $S_z=+1$ 
\\
4
& $eV=+25$
& $\downarrow$
& Strong
& $S_z=-1$ 
\\
\hline\hline
\end{tabular}
\end{table}

\section{Discussion and conclusions\label{sec:conclusions}}
In conclusion, we have demonstrated that molecular crystals formed by combining doped and undoped nanographenes can simultaneously realize half-metallicity and fully compensated ferrimagnetism — two properties long sought in condensed matter physics but rarely~\cite{liu20205FullyCompFerri2D, Kawamura2024organicFerri} found together, and never before in a purely organic, pi-electron system. The nitrogen-substituted [2,A3]triangulene crystal emerges as a singular platform: its Fermi level sits at a spin-polarized flat band, its magnetic order is robust in the range of tens of Kelvin and its vanishing net magnetization eliminates the stray fields that plague conventional spintronic devices. Beyond the half-metallic ground state, the system hosts a quantized anomalous Hall effect driven by Berry curvature concentrated near the $\Gamma$ point, and a magnon spectrum in which most modes are shielded from Landau damping $\textendash$ a combination that is almost paradoxical in a metal. The accessibility of these excitations to inelastic scanning tunneling spectroscopy, and the robustness of the phenomenology to different dopants, suggests that experimental realization is within reach using existing on-surface synthesis techniques. Taken together, these results establish triangulene crystals as a compelling new frontier for carbon-based spintronics and magnonics, and open a design pathway toward functional organic quantum materials engineered atom by atom.

\begin{acknowledgments}
J.F.-R., J.C.G.H., J.P. and A.T.C.  acknowledge financial support from 
 SNF Sinergia (Grant Pimag,  CRSII5\_205987)
J.F.-R. acknowledges financial funding from 
and
MICIN-Spain (Grants No.  PID2022-141712NB-C22) 
The authors thankfully acknowledge RES resources provided by Barcelona Supercomputing Centre in MareNostrum 5 to FI-2025-3-0039.
The authors acknowledge the use of computer time at MareNostrum 5 (Barcelona Supercomputing Centre) provided through FCT grant 2025.00090.CPCA.A2.

\end{acknowledgments}

\bibliography{bib}

\clearpage
\onecolumngrid

\section*{Supplementary Material: Prediction of Two-Dimensional  $\pi$-Electron Half-Metallic Ferrimagnets}

\author{J. Phillips }
\affiliation{International Iberian Nanotechnology Laboratory (INL), Av. Mestre Jos\'e Veiga, 4715-330 Braga, Portugal}
\email{jan.phillips@inl.int}

\author{J. C. G. Henriques}
\affiliation{International Iberian Nanotechnology Laboratory (INL), Av. Mestre Jos\'e Veiga, 4715-330 Braga, Portugal}
\affiliation{Universidade de Santiago de Compostela, 15782 Santiago de Compostela, Spain}

\author{J. Fern\'andez-Rossier}
\altaffiliation[On permanent leave from ]{Departamento de F\'isica Aplicada, Universidad de Alicante, 03690 San Vicente del Raspeig, Spain.}
\affiliation{International Iberian Nanotechnology Laboratory (INL), Av. Mestre Jos\'e Veiga, 4715-330 Braga, Portugal}
\email{joaquin.fernandez-rossier@inl.int}

\author{A. T. Costa}
\affiliation{Department of Physics, Center of Physics (CF-UM-UP), University of Minho, Campus of Gualtar, 4710-057, Braga, Portugal}
\affiliation{International Iberian Nanotechnology Laboratory (INL), Av. Mestre Jos\'e Veiga, 4715-330 Braga, Portugal}
\email{antonio.costa@inl.int \redmark{check}}

\maketitle

\subsection*{DFT}
\begin{figure}
    \centering
    \begin{tabular}{cc}
       \includegraphics[width=0.8\linewidth]{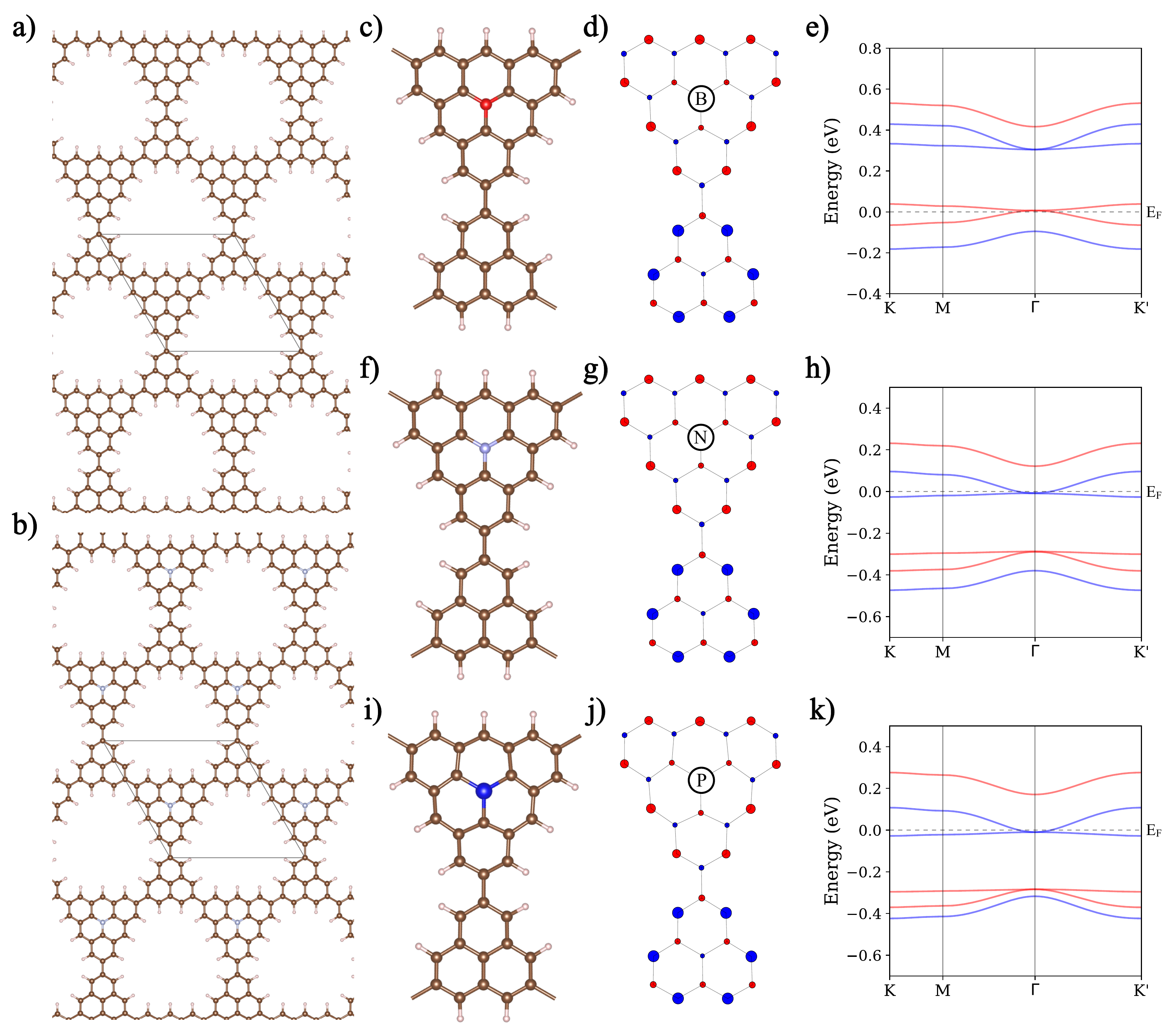} &

    \end{tabular}
    \caption{Crystal structure for a) the undoped [2,3]triangulene and b) the doped triangulene. Brown atoms represent C atoms, light pink represent H atoms and the doped atom at the center of the [3]trianguelne is represented in light blue. Different computed doped structures are represented, when doping with boron c),d),e), with nitrogen f),g),h) and with phosphorus i),j),k). Monomers for the different doped structures are shown c),f),i), together with their respective magnetic profiles d),g),j) and their corresponding bandstructures e),h),k). In the magnetic profiles and the bandstructures, red (blue) represents spin up (down). Fermi level is depicted with a dashed line in the bandstructure plots, and it becomes obvious that for every doped case it lies very close to a nearly flat band.
    }
    \label{fig:doping}
\end{figure}

Density Functional Theory (DFT) results were obtained using the {\sc QUANTUM ESPRESSO} package \cite{QE-2017, QE-2009, doi:10.1063/5.0005082}. We used norm-conserving pseudopotentials\cite{PhysRevLett.43.1494, PhysRevB.43.1993, van_Setten_2018, Hamann2013, garrity2014pseudopotentials, doi:10.1126/science.aad3000}. The plane-wave energy cutoff of was converged to a value of 70 Ry for the wave functions and 280 Ry for the charge density. The Brillouin zone integrals for the relaxations were performed in a converged 15$\times$15$\times$1 k-point grid with a gaussian smearing of 0.001 Ry. In-plane lattice parameterse and internal atomic positions were relaxed for every doped structure (Fig \ref{fig:doping}). The crystals include a converged 9\AA vacuum for the z-axis to avoid inter-molecular interactions.

\begin{figure}
    \centering
    \includegraphics[width=\linewidth]{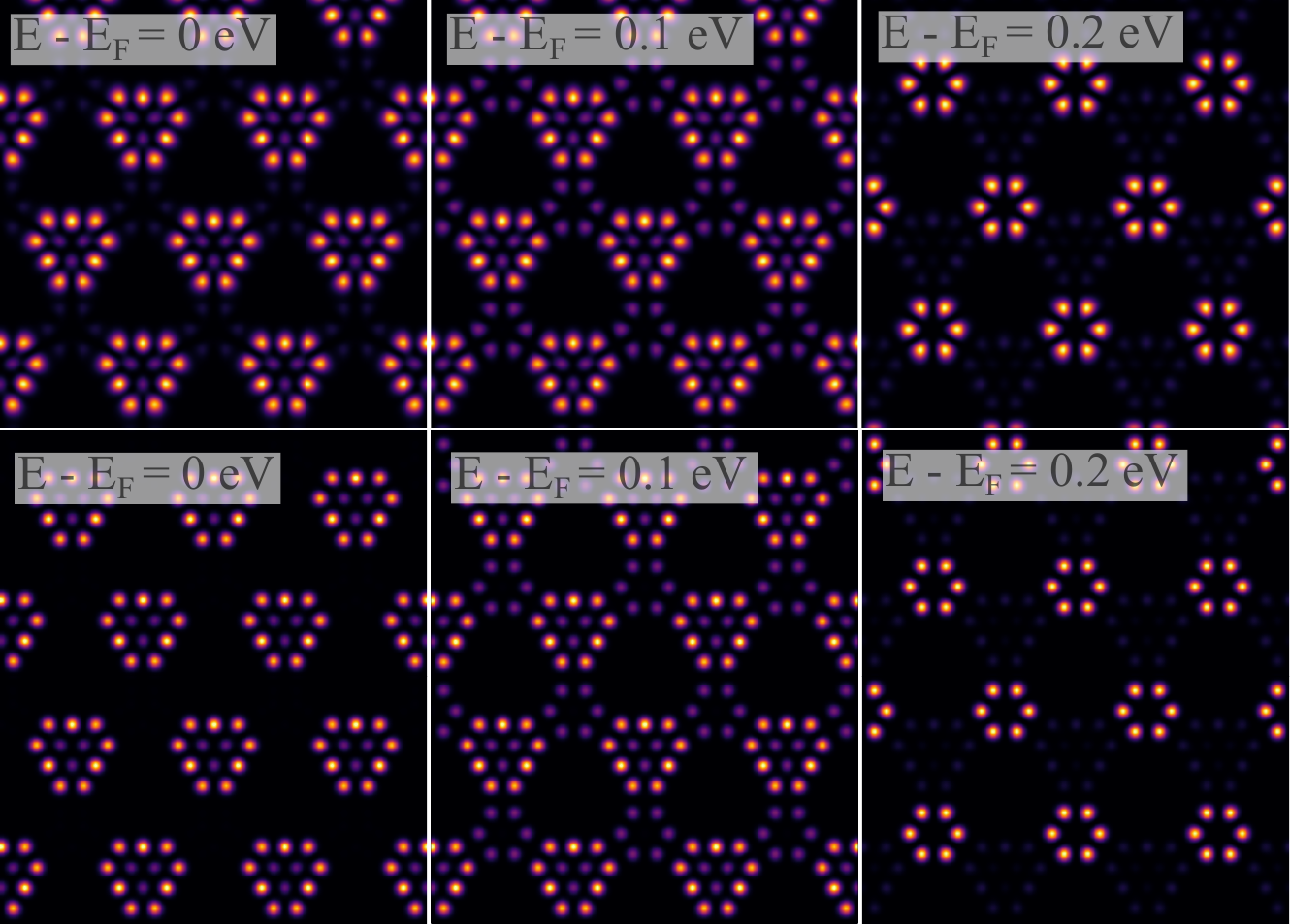}
    \caption{Comparison of the computed LDOS using DFT (top row) and Mean-field Hubbard (bottom row). For every caption the energy at which the LDOS is computed is shown, referenced to the Fermi level.}
    \label{fig:ldos_sup}
\end{figure}

The computed LDOS in Fig. \ref{fig:ldos_sup} shows that the flat band at the Fermi level comes mainly from states of the [3]triangulene. When scanning the band 0.1 eV higher in energy, we observe the contribution from the [2]trianguelene states start to appear. At 0.2 eV the [3]triangulene states disapear in the LDOS and the contribution is solely from the [2]triangulene. These statements are true for both DFT and MFH and is yet another example of the excellent agreement between them in this system.

\subsection*{Magnon spectrum}
The energies and lifetimes of the magnons with $S^z=-1$ and $S^z=1$ were obtained from the dynamic transverse spin susceptibilities, calculated within the random phase approximation (RPA),
\begin{eqnarray}
    \chi^{+-}_{ll'}(t)\equiv -i\theta(t)\left\langle\left[S^+_l(t),S_{l'}^-(0)\right]\right\rangle,\nonumber\\
    \chi^{-+}_{ll'}(t)\equiv -i\theta(t)\left\langle\left[S^-_l(t),S_{l'}^+(0)\right]\right\rangle,
    \label{eq:chidefinition}
\end{eqnarray}
where $S^+_l\equiv a^\dagger_{l\up}a_{l\dn}$, $S^-_l\equiv a^\dagger_{l\dn}a_{l\up}$, $l$ is a site index, $\langle\cdot\rangle$ is a thermodynamic average and $\theta(t)$ is the Heaviside unit step function.
The magnon spectral densities are defined as the imaginary part of the frequency-domain susceptibilities,
\begin{equation}
    \chi^{\perp}_{ll'}(\omega) = \int_{-\infty}^\infty e^{i\omega t} \chi^{\perp}_{ll'}(t)dt,
\end{equation}
where $\perp$ stands for either $+-$ or $-+$.
The RPA consists in decoupling the dynamics of the majority and minority spin electrons at the level of the equation of motion for the susceptibility or, equivalently, summing up the ladder diagrams in the perturbation series to all orders. Both procedures result in the same final expression for the susceptibility \cite{Barbosa2001},
\begin{eqnarray}
    \chi^\perp_{ll'}(\omega) =  \tilde{\chi}^\perp_{ll'}(\omega) + \sum_m \tilde{\chi}^\perp_{lm}(\omega)U_{m}\chi^\perp_{ml'}(\omega),
\end{eqnarray}
where $\tilde{\chi}^\perp$ is the susceptibility defined in Eq.~\ref{eq:chidefinition} calculated for the mean-field Hamiltonian (Eq.~\ref{eq:HubbardMF}). In this work the site indices $l,l'$ can be separated into a unit cell index (indicating the position of a triangulene dimer) and an atom index (indicating the position of an atom inside the unit cell). Due to translation symmetry in the unit cell index, it is possible to adopt a mixed wave-vector/atomic site representation in which the transverse susceptibility is a matrix in atomic site indices $l$, whose elements are functions of a two-dimensional wave vector $\vec{q}$. The RPA equation then becomes, 
\begin{eqnarray}
    \chi^\perp_{ll'}(\omega;\vec{q}) =  \tilde{\chi}^\perp_{ll'}(\omega;\vec{q}) + \sum_m \tilde{\chi}^\perp_{lm}(\omega;\vec{q})U_{m}\chi^\perp_{ml'}(\omega;\vec{q}).
\end{eqnarray}
The mean-field susceptibilities $\tilde{\chi}^\perp_{ll'}(\omega;\vec{q})$ were calculated by numerical evaluation of the $\vec{k}$-space sums,
\begin{equation}
    \tilde{\chi}^{+-}_{lm}(\omega;\vec{q}) = \sum_{\mu,\nu,\vec{k}}\frac{\phi_{l;\up}^{(\mu)*}(\vec{k})\phi_{m;\up}^{(\mu)}(\vec{k})\phi_{l;\dn}^{(\nu)}(\vec{k}+\vec{q)}\phi_{m;\dn}^{(\nu)*}(\vec{k}+\vec{q})\left\{f[E_\mu^\up(\vec{k})]-f[E_\nu^\dn(\vec{k}+\vec{q})]\right\}}{\omega + E_\mu^\up(\vec{k}) - E_\nu^\dn(\vec{k}+\vec{q})+i\eta},
\end{equation}
and an analogous expression for $\tilde{\chi}^{-+}$, with the spin indices reversed. Here $\eta$ is a (small) regularization parameter, and
\begin{equation}
    |\vec{k};\mu;\sigma\rangle \equiv \sum_{l,\vec{R}}e^{i\vec{k}\cdot\vec{R}}\phi_{l;\sigma}^{(\mu)}(\vec{k})|l;\vec{R}\rangle,
\end{equation}
are the eigenstates of the mean-field Hamiltonian $H_\mathrm{MF}^\sigma$
for spin direction $\sigma$, with energies $E_\mu^\sigma(\vec{k})$, written in the basis of atomic-like states $|l;\vec{R}\rangle$ localized at atomic site $l$ in unit cell $\vec{R}$. For the results presented in this work we adopted $\eta=0.1$~meV and summed over 478800 $\vec{k}$ points.

As shown in Fig.~3 of the main text (also reproduced here for ease of reference), the $S^z=1$ magnon dispersion has two branches, that are commonly referred to as acoustic (low-energy, linearly vanishing at $\Gamma$) and optical (high-energy, usually an extreme at $\Gamma$). The energy points for the optical branch around the $\Gamma$-point are conspicuously missing from that plot. Here we address the reason for that absence, which is as follows. The acoustic branch has a well-defined signature with vanishing linewidth over the whole Brillouin zone. The optical branch, however, merges with the Stoner continuum around the $\Gamma$-point, and ceases to be identifiable as a true collective mode. In Fig.~\ref{fig:magnons} we show the evolution of the $S^z=1$ magnon spectral density along the $K-\Gamma$ direction by plotting $A^\gamma(\omega,\vec{q})$, where
\begin{equation}
    A(\omega,\vec{q})\equiv -\frac{1}{\pi}\mathrm{Im}\sum_l\chi^{-+}_{ll}(\omega,\vec{q}),
\end{equation}
$\chi^{-+}_{ll}(\omega,\vec{q})$ is the transverse spin susceptibility associated with $S^z=1$ magnons, $l$ are sites in the crystal's unit cell, and $\gamma=0.2$ is a compression factor to help the visualization of the very sharp peaks in the spectral function. The optical mode starts as a well-defined collective excitation at the $K-$point, but as its energy increases it quickly approaches the Stoner continuum, merging with it and becoming overdamped at a point between 70\% and 80\% of the way from $K$ to $\Gamma$.

\begin{figure}
    \includegraphics[width=0.98\linewidth]{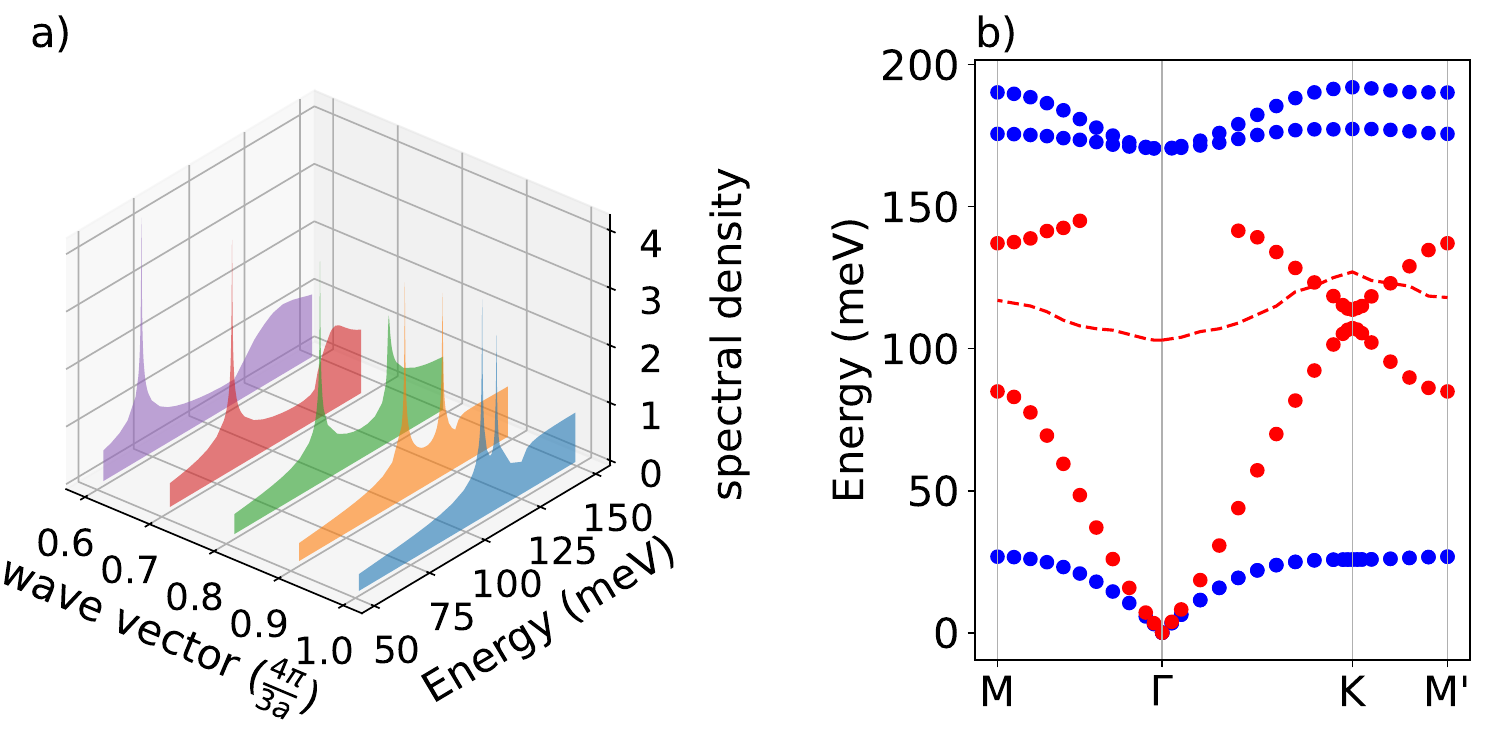}
    \caption{a) Power-compressed spectral density $A^\gamma(\omega,\vec{q})$ as a function of magnon energy $\hbar\omega$ for selected wave vectors along $\Gamma-K$. As the wave vector departs from $K$ (blue spectral density) the energy of the acoustic magnon (lower-energy peak) 
    decreases, whereas that of the optical magnon (higher-energy peak) increases,
    eventually leading to the disappearance of the optical magnon peak into the Stoner continuum. b) Magnon energies for selected wave vectors extracted from the fermionic model, for the two magnon flavors $S^z=-1$ (blue) and $S^z=1$ (red). The red dashed line marks the lower energy boundary of the $S^z=1$ Stoner continuum. }
    \label{fig:magnons}
\end{figure}


\end{document}